\pdfoutput=1
\documentclass[
aip,
superscriptaddress, 
amsmath,amssymb,preprint,nofootinbib]{revtex4-1}

 \usepackage{graphicx}
 \graphicspath{{pics/}} 
\usepackage{dcolumn}
\usepackage{bm}

\usepackage[utf8]{inputenc}
\usepackage[T1]{fontenc}
\usepackage{newtxmath} 
\usepackage{etoolbox} 
\usepackage{mathtools}

\usepackage[retainorgcmds]{IEEEtrantools} 

\usepackage{extarrows} 

\usepackage[singlelinecheck=false,justification=raggedright,bf]{caption}
\usepackage{subcaption}

\newcommand{\secref}[1]{Sec.~\ref{#1}}
\newcommand{\appref}[1]{App.~\ref{#1}}
\newcommand{\Eqref}[1]{Eq.~(\ref{#1})}

\newcommand{\figref}[1]{Fig.~\ref{#1}}
\newcommand{\tabref}[1]{table~\ref{#1}}

\usepackage{xcolor}
\usepackage{tikz}

\usepackage{ytableau}
\ytableausetup{mathmode, centertableaux, boxsize=1.5ex}

\usepackage{cancel}

\usepackage{pbox}

\usepackage[pdftex,colorlinks]{hyperref} 
 \hypersetup{
     colorlinks,
     linkcolor={red!50!blue},
     citecolor={blue!50!green},
     urlcolor={blue!80!black}
   }

\usepackage[ntheorem]{empheq}
\usepackage[hyperref]{ntheorem}

\newcommand{\qed}{\hfill\tikz{\draw[draw=black,line width=0.6pt] (0,0)
    rectangle (2.8mm,2.8mm);}\bigskip}

\makeatletter
\newcommand{\vast}{\bBigg@{3}}
\newcommand{\Vast}{\bBigg@{4}}
\newcommand{\Vvast}{\bBigg@{5}}
\newcommand{\VAST}{\bBigg@{6}}
\makeatother

\usepackage[nameinlink]{cleveref} 

\makeatletter
\def\@email#1#2{
 \endgroup
 \patchcmd{\titleblock@produce}
  {\frontmatter@RRAPformat}
  {\frontmatter@RRAPformat{\produce@RRAP{*#1\href{mailto:#2}{#2}}}\frontmatter@RRAPformat}
  {}{}
}
\makeatother

\begin{document}

\newcommand{\RPic}[2][{}]{\hspace{-0.4mm}\pbox{\textwidth}{\includegraphics[#1]{{#2}}}\hspace{-0.4mm}}
\newcommand{\FPic}[2][{}]{\hspace{-0.27mm}\pbox{\textwidth}{\hspace*{.5ex}\includegraphics[#1]{{#2}}\hspace*{.5ex}}\hspace{-0.27mm}}
\newcommand{\ybox}[1]{\begin{ytableau} #1 \end{ytableau}}

\newcommand{\SUN}{\mathsf{SU}(N)}
\newcommand{\suN}{\mathsf{su}(N)}
\newcommand{\SU}[1]{\mathsf{SU}(#1)}
\newcommand{\GLN}{\mathsf{GL}(N)}
\newcommand{\MixedPow}[2]{V^{\otimes
    #1}\otimes\left(V^*\right)^{\otimes #2}}
\newcommand{\Pow}[1]{V^{\otimes #1}}

\preprint{MCnet-23-24}
\title{Wigner $6j$ symbols with gluon lines:\\completing the set of 6j symbols required for color decomposition}

\author{Stefan Keppeler}
\affiliation{Fachbereich Mathematik, Universit\"at T\"ubingen, Auf der Morgenstelle 10, 72076 T\"ubingen, Germany}

\author{Simon Pl\"atzer}
\affiliation{Institute of Physics, NAWI Graz, University of Graz, Universit\"atsplatz 5, A-8010 Graz, Austria}
\affiliation{Particle Physics, Faculty of Physics, University of Vienna, Boltzmanngasse 5, A-1090 Wien, Austria}
\affiliation{Erwin Schr\"odinger Institute for Mathematics and Physics, University of Vienna, Boltzmanngasse 9, A-1090 Wien, Austria}

\author{Malin Sjodahl}
\affiliation{Department of Physics, Lund University, Box 118, 221 00 Lund, Sweden}

\date{\today}

\begin{abstract}

We construct a set of Wigner $6j$ symbols with gluon lines (adjoint
  representations) in closed form, expressed in terms of similar $6j$
  symbols with quark lines (fundamental representations). Together with
  Wigner $6j$ symbols with quark lines, this gives a set of $6j$
  symbols sufficient for treating QCD color structure for any number
  of external particles, in or beyond perturbation theory.
  This facilitates a complete treatment of QCD color structure in terms of
  orthogonal multiplet bases, without the need of ever explicitly
  constructing the corresponding bases.
  We thereby open up for a completely representation theory based treatment of SU(N) color structure, with the potential of significantly speeding up the color structure treatment.

\end{abstract}

\maketitle

\section{Introduction}
\label{sec:intro}

A major challenge for accurate predictions of collision rates for
processes involving many colored patrons, is the treatment of the $\SUN$
color space associated with QCD.  This challenge is typically
addressed by expanding in color bases, often trace bases
\cite{Paton:1969je, Berends:1987cv, Mangano:1987xk, Mangano:1988kk,
  Kosower:1988kh, Nagy:2007ty, Sjodahl:2009wx, Alwall:2011uj,
  Sjodahl:2014opa, Platzer:2012np,Platzer:2018pmd} or color-flow bases
\cite{tHooft:1973alw,Kanaki:2000ms,Maltoni:2002mq,Platzer:2013fha,AngelesMartinez:2018cfz,DeAngelis:2020rvq,Platzer:2020lbr},
and sometimes accompanied by sampling of the color states
\cite{Platzer:2013fha,DeAngelis:2020rvq,Platzer:2020lbr,Isaacson:2018zdi}.

A problem with the trace and color-flow bases is that they are only
orthogonal in the limit $N\to \infty$, and in fact overcomplete for
many particles; for high multiplicities they are severely overcomplete
\cite{Sjodahl:2015qoa}, with a dimension that scales as the factorial
of the number of gluons plus quark-antiquark pairs.  If one does not
want to exploit sampling over different color structures\footnote{Typically
the sampling is also accompanied by detailed relations of color flows and kinematic quantities.},
like done in for example the CVolver program
\cite{Platzer:2013fha,Forshaw:2021mtj,DeAngelis:2020rvq}, this gives
rise to a major bottle neck for the squaring of the color structure,
which then scales as a factorial square.

It appears appealing to explore minimal orthogonal bases. This is
accomplished by multiplet bases \cite{Kyrieleis:2005dt,
  Dokshitzer:2005ig,Sjodahl:2008fz,Beneke:2009rj,Keppeler:2012ih,
  Du:2015apa,Sjodahl:2015qoa,Keppeler:2013yla,Alcock-Zeilinger:2016bss,
  Alcock-Zeilinger:2016sxc,Alcock-Zeilinger:2016cva,Sjodahl:2018cca},
which rely on the Clebsch-Gordan decomposition of the
involved particle representation for constructing orthogonal
bases. Examples of
multiplet bases can be found in Refs.~\citenum{Kyrieleis:2005dt,
  Dokshitzer:2005ig,Sjodahl:2008fz,Beneke:2009rj}, and a general
construction in
Refs.~\citenum{Keppeler:2012ih,Alcock-Zeilinger:2016cva}.

However, it is possible to do better than that: Any color structure
can be decomposed into a multiplet basis \textit{without} explicitly
constructing this basis, by making use of the group invariant Wigner
$6j$ symbols (here $6j$s for short, also known as $6j$ coefficients,
Racah coefficients, or Racah $W$ coefficients, up to signs), along
with Wigner $3j$ coefficients and dimensions of representations.  The
problem of decomposing the color structure is then essentially
\textit{reduced} to the problem of finding a sufficient set of $6j$
symbols for the color decomposition in question. Some work in this
direction has been pursued in Refs. \citenum{Sjodahl:2015qoa,Sjodahl:2018cca}, where
symmetry is exploited to recursively calculate a set of $6j$ symbols
applicable for processes with a limited number of partons. Other
recent work obtains $\SU{3}$ $6j$ symbols numerically, by first
calculating $\SU{3}$ Clebsch-Gordan
coefficients\cite{Alex_2011,Dytrych:2021qwe}.  For $\SU{2}$, the
problem is addressed in Ref.~\citenum{Johansson:2016}.

In a recent paper \cite{Alcock-Zeilinger:2022hrk}, we started to
explore a third avenue, namely to recursively derive $6j$ symbols in
terms of other $6j$ symbols and dimensions of representations. We
there derived closed forms of a set of $6j$ symbols characterized by
having quarks (fundamental representations) in opposing
positions.  In the present paper, we complete this set of $6j$ symbols
with symbols where two of the 
    opposing representations are quarks \textit{or} gluons (adjoint representations), and $6j$s
    where one vertex only contains fundamental and adjoint 
    representations,
  whereas the other representations are arbitrary. As we will see,
  this class of $6j$s define a complete set for decomposing any color
  structure appearing in the standard model.

We lay out the basics
of $\SUN$ color calculations using the birdtrack method
  in \secref{sec:computational}. In
\secref{sec:gluon6js-4cases} we go through a general procedure for
decomposing the color structure. This allows us to identify a
set of $6j$ symbols that is
sufficient  to decompose any color structure to any order in perturbation
theory.  While one of the necessary classes of $6j$ symbols is
calculated in Ref.~\citenum{Alcock-Zeilinger:2022hrk}, the remaining
ones are calculated in \secref{sec:closed-form-expressions}, after a
careful discussion on how to define vertices in \secref{sec:vertices}.
Finally, we make concluding remarks in \secref{sec:conclusions}.

\section{Reducing ${\SUN}$ color structure in birdtrack notation}
\label{sec:computational}

In this section we briefly outline how to calculate $\SUN$ invariants,
using the birdtrack method, and assuming knowledge of a sufficient set
of Wigner $3j$\footnote{We will later normalize the $3j$s to 1.} and Wigner
$6j$~symbols.  It is worth remarking that while our discussion focuses
on $\SUN$, in particular $\SU{3}$, this reduction
method is applicable for any Lie group. For a full, comprehensive
introduction to the birdtrack formalism, we refer to
Ref.~\citenum{Cvi08}, a minimal introduction can be found in
Appendix~A of Ref.~\citenum{Keppeler:2012ih}, whereas a more
pedagogical account is written up in
Ref.~\citenum{Keppeler:2017kwt}. Examples of birdtrack calculations
for QCD can be found in Refs.~\citenum{Du:2015apa} and
\citenum{Sjodahl:2015qoa}.

\begin{figure}[t]
  \raisebox{-1.2 cm}{\includegraphics[scale=0.5]{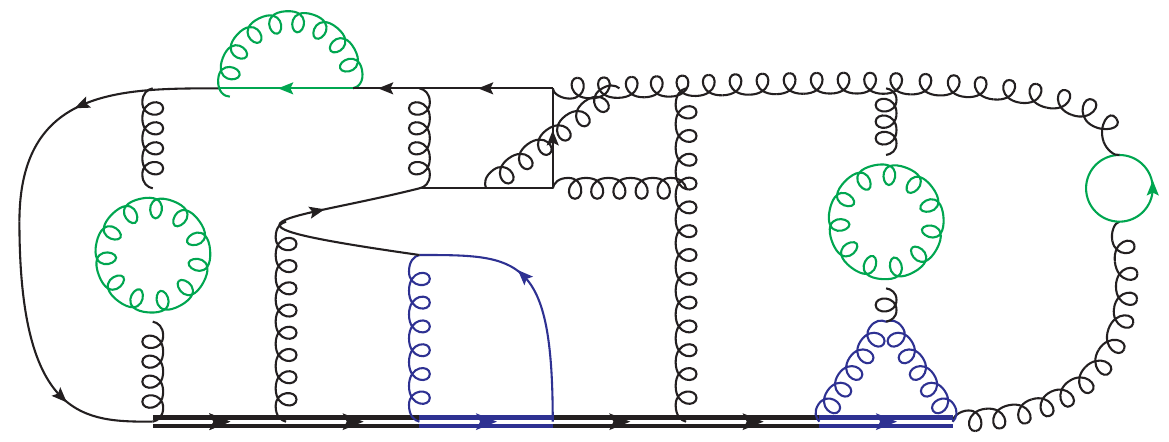}}
  \caption{Example of a fully contracted color structure with triplet and
    and adjoint representations, contracted with a set of general representations,
    denoted by double lines. These may come from a multiple basis vector,
    or arise previous contraction steps. Note that ``bubbles'', in green,
    can be contracted away using \Eqref{eq:self-energy}, and that vertex
    corrections (in blue) can be removed using \Eqref{eq:vertex-correction}.
    To address the remaining color structure, the completeness relation
    \Eqref{eq:completeness-relation} is in general needed.
    \label{fig:full contraction}
  }
\end{figure}

As we are interested in fully color summed (averaged) color
structures, every color structure can be seen as a fully connected
graph of $\SUN$ representations, for example as in \figref{fig:full
  contraction}.  This entails of course triplet and octet
representations but also the higher dimensional irreducible
representations (irreps), used in the construction of multiplet bases,
or appearing during the calculations.  In the end, we want to
calculate a scalar product in color space, for example between a
Feynman diagram and basis vector in an orthogonal multiplet basis.

Generally, the color structure then consists of a fully connected graph.
The graph contains loops of various length, for example, we may encounter
\begin{equation}
  \label{eq:6VertexLoopOnly}
    \raisebox{-1.2 cm}{\includegraphics[scale=0.5]{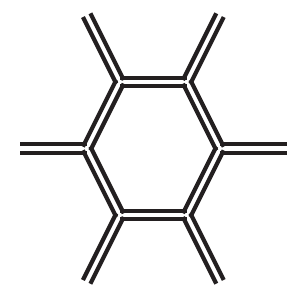}
  }\;,
\end{equation}
where the double lines denote any irrep of $\SUN$, and in general should
be supplied with representation labels and arrow directions, which we
suppress here for readability.

While short loops of length up to three can be immediately removed (see below),
the fall-back method to handle long loops is to split them up
to shorter loops by repeated insertion of the completeness relation
\begin{equation}
  \label{eq:completeness-relation}
  \RPic{GenRep-beta-gamma}
  \; =
  \sum_{\delta}
  \frac{d_{\delta}}{\RPic{3j-gammaSTAR-beta-delta}}
  \;
  \RPic{GenRep-gammaVbeta-delta-ProjOps}
  \ ,
\end{equation} 
where $d_{\delta}$ denotes the dimension of the irrep $\delta$,
appearing in the Clebsch-Gordan decomposition, and where the
denominator is a Wigner $3j$ symbol.

Tracing both sides of this equation, and using 
\begin{equation}
  \label{eq:rep-loop}
  \raisebox{-0.5 cm}{\includegraphics[scale=0.5]{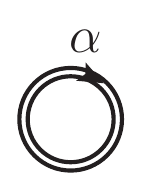}} 
  =d_\alpha\;,  
\end{equation} 
it is clear that the completeness relation implies
$d_\beta d_\gamma =\sum_\delta d_\delta$, as anticipated.

Applying the completeness relation~\eqref{eq:completeness-relation} to
two of the representations in \Eqref{eq:6VertexLoopOnly}, marked in
red below, schematically results in
\begin{equation}
  \label{eq:loop-reduced}
\raisebox{-0.45\height}{
	\includegraphics[scale=0.5]{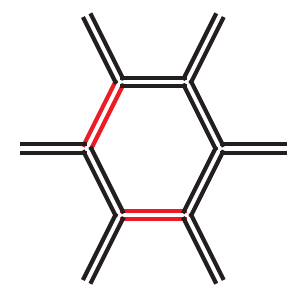}
}
\xlongequal{\eqref{eq:completeness-relation}}
\sum_{\alpha}{
\frac{
	d_\alpha
}{
	\hspace{-2mm}
	\raisebox{-0.45\height}{
		\includegraphics[scale=0.3]{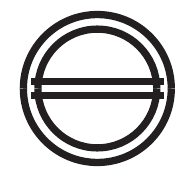}
	}
	\hspace{-2mm}
}
\raisebox{-0.45\height}{
	\includegraphics[scale=0.5]{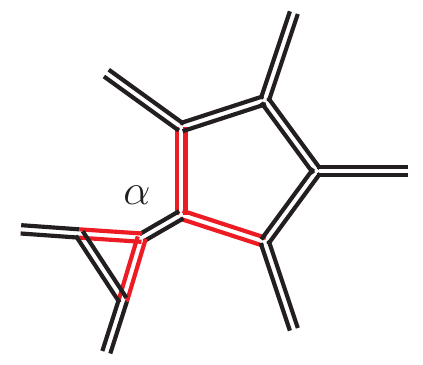}
}
}\;,
\end{equation}
where we now have a ``vertex correction'' loop with three internal
representations. This loop can be removed using the Wigner $6j$ symbols
\begin{equation}
  \label{eq:vertex-correction}
  \RPic{GenRep-alpha-sigma-rhoSTAR--VCorr-delta-gamma-betaSTAR}
  \; =
     \sum_a
     \;
     \frac{1}{\FPic{3j-sigmaSTAR-alpha-rho-VertexLabelsVa}}
     \;
     \underbrace{
       \RPic{6j-delta-rho-sigma--beta-gamma-alpha--CornerLabelsV3a}
     }_{\text{Wigner-$6j$}}
     \hspace{2mm}
     \RPic{Vertex-alpha-sigma-rhoSTAR--V--VLabela}
     \ .
\end{equation}
The sum above runs over instances $a$ of the irrep $\rho$ in $\alpha
\otimes \sigma$, for example the two octets in $8\otimes 8=1\oplus8\oplus 8\oplus 10 \oplus \overline{10}\oplus 27$.  In this paper, every encountered vertex will contain
at least one fundamental or adjoint representation, implying that most
often there is only one instance, but for $A \otimes \sigma$, with $A$
being the adjoint representation for $\SUN$, and $\sigma$ being an
arbitrary irrep, there can be up to $N-1$ representations of type
$\sigma$ \cite{Keppeler:2012ih}. We will choose the corresponding
vertices to be mutually orthogonal, in the sense that
\begin{equation}
  \label{eq:scalar-product-and-3j}
  \scalebox{1.5}{\Bigg\langle}
  \raisebox{-0.45\height}{
    \includegraphics[scale=0.5]{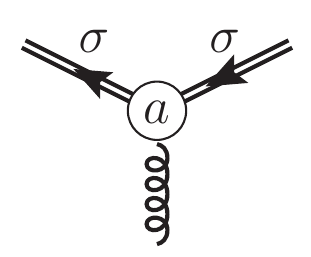}}
  \ , \
   \raisebox{-0.45\height}{\includegraphics[scale=0.5]{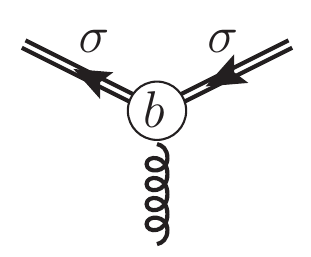}}
   \scalebox{1.5}{\Bigg\rangle}
   = \raisebox{-0.45\height}{\includegraphics[scale=0.5]{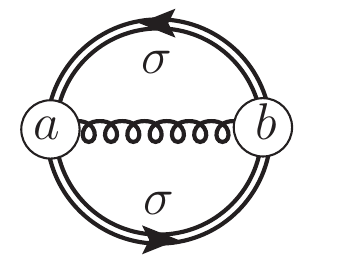}}
  \stackrel{!}{=} 0 \qquad \text{if} \qquad a\ne b
  \, .
\end{equation}

Furthermore, for the $6j$ symbols that we derive, we choose to
normalize our vertices such that
\begin{equation}
  \label{eq:3jnorm}
   \raisebox{-0.45\height}{\includegraphics[scale=0.5]{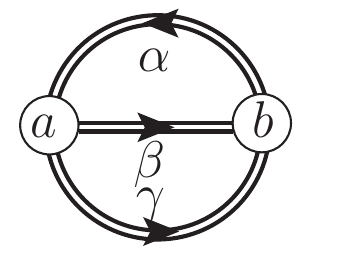}}\equiv \delta_{ab}\;,\quad \text{for all non-vanishing vertices,}
\end{equation}
i.e., the $3j$ coefficients are normalized to one.  This implies in
particular that
\mbox{$\raisebox{-0.6cm}{\includegraphics[scale=0.5]{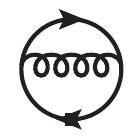}}=
  1$}, in contrast to the standard QCD normalization
\mbox{$\raisebox{-0.6cm}{\includegraphics[scale=0.5]{quark-gluon-3j}}=
  \frac{1}{2}(N^2-1)$},
for the generator normalization
$\text{tr}[t^at^b]=\frac{1}{2}\delta^{a b}$. We explain in
\appref{app:restore-3j} how to easily transform our results to any
desired normalization.

Note that after having applied \Eqref{eq:vertex-correction} to
\Eqref{eq:loop-reduced}, we are left with a
loop with one representation less,
  \begin{equation}\label{eq:LoopContraction}
\raisebox{-0.45\height}{
	\includegraphics[scale=0.47]{6VertexLoop}
}
\xlongequal{\eqref{eq:completeness-relation}}
\sum_{\alpha}{
\frac{
	d_\alpha
}{
	\hspace{-2mm}
	\raisebox{-0.45\height}{
		\includegraphics[scale=0.3]{Wig3jNoRepLabels}
	}
	\hspace{-2mm}
}
\raisebox{-0.45\height}{
	\includegraphics[scale=0.47]{5VertexLoopBeforeSchur}
}
}
\xlongequal{\eqref{eq:vertex-correction}}
\sum_{\alpha}{
\frac{
	d_\alpha
}{
	\hspace{-2mm}
	\raisebox{-0.45\height}{
		\includegraphics[scale=0.3]{Wig3jNoRepLabels}

	}
	\hspace{-2mm}
}
\frac{
	\hspace{-1mm}
	\raisebox{-0.2\height}{
		\includegraphics[scale=0.4]{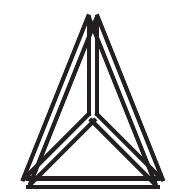}
	}
}{
	\hspace{-2mm}
	\raisebox{-0.45\height}{
		\includegraphics[scale=0.3]{Wig3jNoRepLabels}
	}
	\hspace{-2mm}
}
\raisebox{-0.45\height}{
	\includegraphics[scale=0.47]{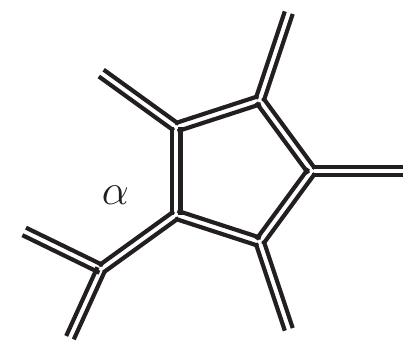}
}
}.
\end{equation}

Repeatedly applying this procedure 
to \Eqref{eq:6VertexLoopOnly}, it is thus
possible to reduce loops with any number of internal representations
down to loops of length three (removed using \Eqref{eq:vertex-correction}) or length two, removed using
the ``self energy'' relation
\begin{equation}
  \label{eq:self-energy}
  \raisebox{-0.8 cm}{\includegraphics[scale=0.5]{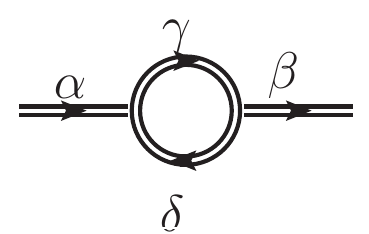}} 
  = \frac{\includegraphics[scale=0.5]{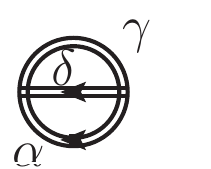}}{d_\alpha}
  \raisebox{-0.1 cm}{\includegraphics[scale=0.5]{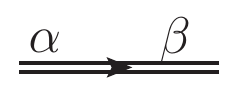}}
  \ .
\end{equation} 

In this way, assuming the knowledge of the $6j$ symbols,
it is possible to
reduce any fully connected graph to a number. 

It should be noted that the required set of $6j$s depends
on how the contraction is performed, and what basis vectors
are used. In the present paper, we consider basis vectors
of the form,
\begin{equation}
  \label{eq:basis vector}
  \raisebox{-0.45\height}{\includegraphics[scale=0.5]{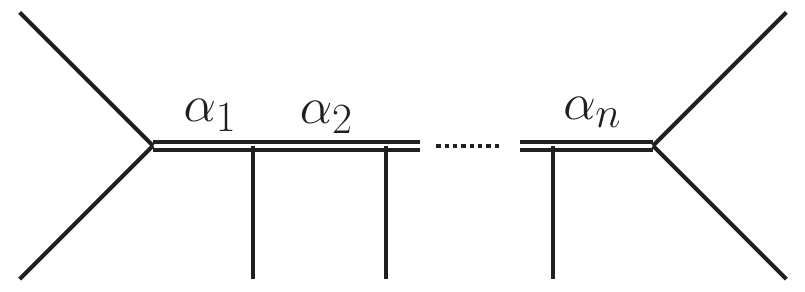}}\;,
\end{equation}
where we have a backbone chain of general representations
$\alpha_1,\alpha_2,\cdots, \alpha_n$, denoted by double lines with
suppressed arrows, to which the external particle representations,
octets, triplets and antitriples, denoted by single lines, are
attached (also with suppressed arrows).
To the authors knowledge all multiplet bases in the literature are of
this form.
One could also imagine bases
where general irreps are contracted in vertices. Such bases will
require $6j$ symbols beyond those presented here, and are therefore
beyond the scope of this paper. We emphasize once more, that the
philosophy underlying the present work is to avoid explicitly
constructing any bases, and instead achieve a decomposition using $6j$s.

\section{A sufficient set of \texorpdfstring{$6j$}{6j}~symbols for decomposing color structure}

\label{sec:gluon6js-4cases}

In this section we identify a sufficient set of $6j$ symbols for
decomposing color structure into the orthogonal basis vectors in
\Eqref{eq:basis vector} to any order in perturbation theory.  We start
out with considering tree-level color structures, and return to higher
orders later.

Again, letting single lines schematically denote triplet, octet or
singlet representations (i.e. representations of the external
particles) and letting double lines denote the general irreps encountered in the
basis vectors, the fully connected graph for a tree-level
color structure contracted with a basis vector, will always contain at least two loops of the form
\begin{equation}\label{eq:Loop1}
\raisebox{-0.4\height}{
  \includegraphics[scale=0.45]{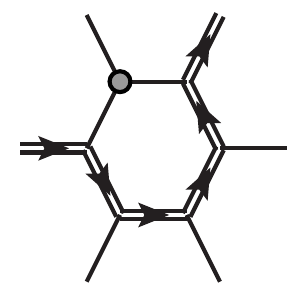}},
\end{equation}
where the characterizing feature is that there is only one vertex from
the initial color structure (the gray blob, representing a quark-gluon
or triple-gluon vertex).  Typically the color structure will contain
many color structures that are trivial to contract using
\Eqref{eq:vertex-correction} and \Eqref{eq:self-energy} , but we here
consider a worst, general case.  To reduce loops of this type, the
completeness relation, \Eqref{eq:completeness-relation}, and the
vertex correction relation, \Eqref{eq:vertex-correction}, can be
applied to the two red representations below
\begin{equation}\label{eq:CRApplicationLoopType1}
  \raisebox{-0.4\height}{
    \includegraphics[scale=0.45]{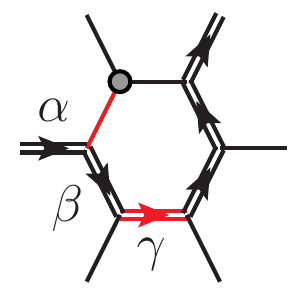}
  }
  =
  \sum_{\psi}{
    \frac{d_{\psi}}{
      \hspace{0.5mm}
      \raisebox{-0.45\height}{
        \includegraphics[scale=0.4]{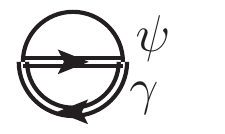}
                
      }
      \hspace{-3mm}
    }
    \raisebox{-0.4\height}{
      \includegraphics[scale=0.45]{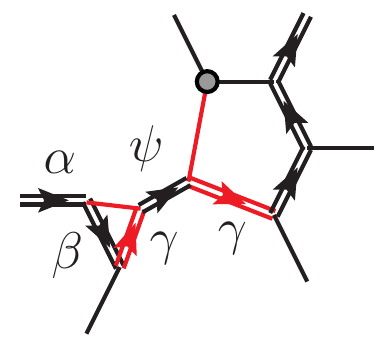}
    }
  }
=
\sum_{\psi,a}{
		\frac{d_{\psi}}{
		\hspace{0.5mm}
		\raisebox{-0.45\height}{
                  \includegraphics[scale=0.4]{Figures/Wig3jCR}
		}
		\hspace{-3mm}
	}
	\frac{
		\hspace{-1mm}
		\raisebox{-0.1\height}{
                  \includegraphics[scale=0.4]{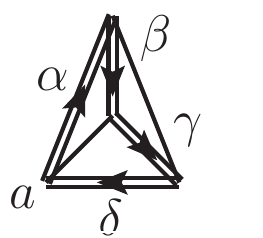}
		}
		\hspace{-5mm}
	}{
		\hspace{0.5mm}
		\raisebox{-0.45\height}{
                  \includegraphics[scale=0.4]{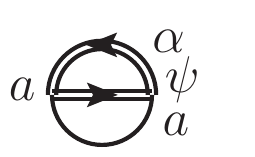}
		}
		\hspace{-4mm}
	}
\raisebox{-0.4\height}{
  \includegraphics[scale=0.45]{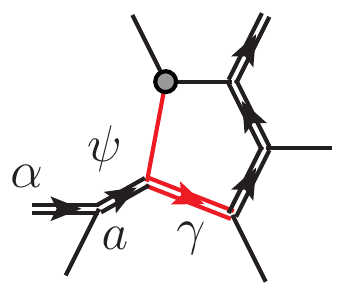}
}
}
.
\end{equation}
Repeating this procedure will eventually result in a vertex correction 
containing the gray blob. (For the loop in the above example, this 
step would need to be repeated two more times.) The vertex correction with the gray blob gives
\begin{equation}\label{eq:CRApplicationLoopType1LastCR}
\raisebox{-0.49\height}{
  \includegraphics[scale=0.4]{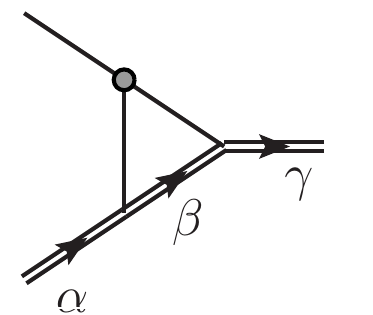}
}
\hspace{-3mm}
=
\sum_{a}{
\frac{
		\hspace{-0.5mm}
		\raisebox{-0.1\height}{
                  \includegraphics[scale=0.4]{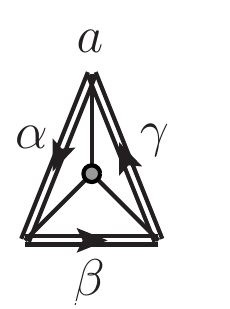}
		}
		\hspace{-4mm}
	}{
		\hspace{-0.5mm}
		\raisebox{-0.45\height}{
                  \includegraphics[scale=0.4]{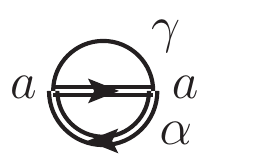}
		}
		\hspace{-4mm}
}
\raisebox{-0.5\height}{
  \includegraphics[scale=0.4]{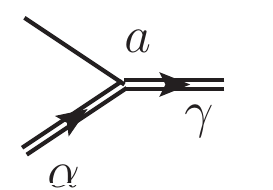}
}
}
\hspace{-3mm}
,
\end{equation}
for some representations $\alpha$, $\beta$ and $\gamma$.
This last step removes two vertices, one gray blob, i.e., a vertex from the initial color structure and one vertex between arbitrary representations in the basis vector, \Eqref{eq:basis vector}. 
As every contracted loop removes one vertex from the basis vector
and one from the color structure to be decomposed,
the resulting graph is topologically equivalent to a graph 
for a tree-level color structure with one less external patron.
After a loop of the form of \Eqref{eq:Loop1} has been contracted, there must
thus exist at least two loops of the type in \Eqref{eq:Loop1} in the
resulting color structure by the above argument. Hence any tree-level
color structure can be completely contracted by repeatedly contracting
loops of the form of \Eqref{eq:Loop1}.

Only treating loops of the form of \Eqref{eq:Loop1} 
is thus sufficient for tree-level color structures.
We now address the situation where the color structure to be
decomposed itself contains loops.
It is then not always possible to choose color loops of the form in
\Eqref{eq:basis vector}. At one-loop this happens
for diagrams where all external partons form a single loop
\begin{equation}\label{eq:NLOAllLoop}
  \raisebox{-0.4\height}{
    \includegraphics[scale=0.4]{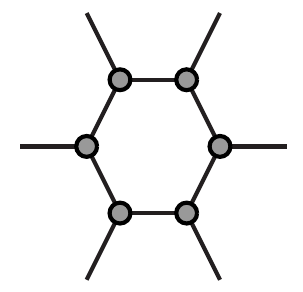}
  }.
\end{equation}
(For all other one-loop color structures there is at least one vertex
with two uncontracted indices, implying that a loop of the from
in \Eqref{eq:Loop1} can be found, such that it is possible to contract
loops as in \Eqref{eq:CRApplicationLoopType1}.)
For color structures of the form of \Eqref{eq:NLOAllLoop}, there always exists loops of the form
\begin{equation}\label{eq:Loop3}
\raisebox{-0.4\height}{
  \includegraphics[scale=0.5]{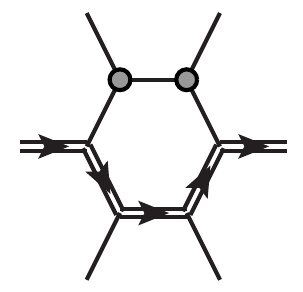}
}.
\end{equation}
Similarly to the loop in \Eqref{eq:Loop1}, the steps detailed in 
\Eqref{eq:CRApplicationLoopType1} remain valid. However, at the end,
instead of contracting a loop of the form in \Eqref{eq:CRApplicationLoopType1LastCR},
a loop with four vertices is encountered,
\begin{eqnarray}
\label{eq:CRApplicationLoopType3}
\raisebox{-0.5\height}{
  \includegraphics[scale=0.4]{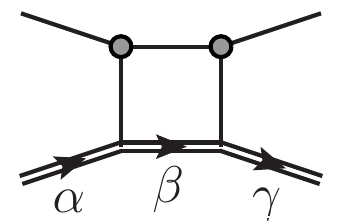}
}
\hspace{-3mm}
&=&
\sum_{\psi}{
\frac{d_\psi}{
	\hspace{-0.5mm}
	\raisebox{-0.45\height}{
          \includegraphics[scale=0.4]{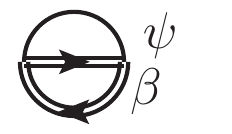}
	}
	\hspace{-4mm}
}
\raisebox{-0.5\height}{
  \includegraphics[scale=0.4]{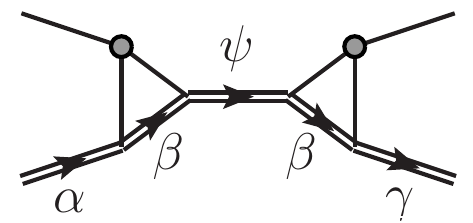}
}
}
\hspace{-3mm}
\nonumber\\
&=&
\sum_{\psi,a,b}{
\frac{d_\psi}{
	\hspace{-0.5mm}
	\raisebox{-0.45\height}{
		\includegraphics[scale=0.4]{Figures/Wig3j_CR}
	}
	\hspace{-4mm}
}
\frac{
		\hspace{-0.5mm}
		\raisebox{-0.1\height}{
                  \includegraphics[scale=0.4]{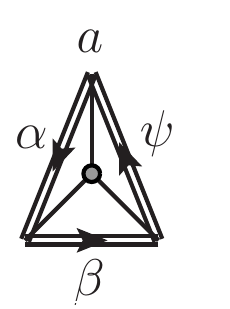}
		}
		\hspace{-4mm}
	}{
		\hspace{-1mm}
		\raisebox{-0.45\height}{
                  \includegraphics[scale=0.4]{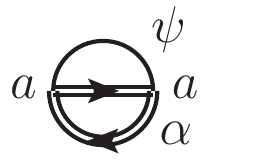}
		}
		\hspace{-4mm}
}
\frac{
		\hspace{-0.5mm}
		\raisebox{-0.1\height}{
                  \includegraphics[scale=0.4]{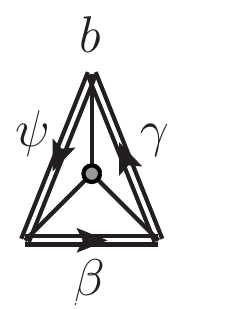}
		}
		\hspace{-4mm}
	}{
		\hspace{-1mm}
		\raisebox{-0.45\height}{
                  \includegraphics[scale=0.4]{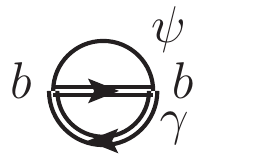}
		}
		\hspace{-4mm}
}
\raisebox{-0.49\height}{
  \includegraphics[scale=0.4]{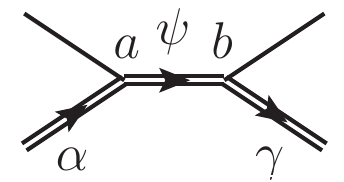}
}
}
\hspace{-1.5mm}
.
\end{eqnarray}
Treating a loop like this removes two vertices from the color
structure, and none from the basis vector. Since a
  one-loop color structure has two vertices more than a tree-level
  color structure for the same process, the
number of vertices after the contraction matches a tree-level
structure. Since two legs which initially belonged to the loop of the
color structure to be decomposed (the upper legs in
\Eqref{eq:CRApplicationLoopType3}) now attach directly to the sequence
of basis vector representations, the topology after the contraction is
equivalent to that of a tree-level color structure.  Note that a loop
of the type in \Eqref{eq:Loop3}, need not be the first loop to be
contracted (most one-loop diagrams contain no loop of the form in
\Eqref{eq:NLOAllLoop}), but such loops may at some point be
encountered, and necessary to contract to continue the reduction.
In this way, we can thus contract any one-loop diagram.  Color
structures of arbitrary order in perturbation theory can be decomposed
by contracting loops similar to \Eqref{eq:CRApplicationLoopType1} and \Eqref{eq:CRApplicationLoopType3},
possibly with more than two (cf. \Eqref{eq:CRApplicationLoopType3})
vertices from the initial color structure. The final steps in the contraction would then
proceed as in \Eqref{eq:CRApplicationLoopType3}, but with more completeness relations inserted.

In the above color decomposition procedure, we can identify
a minimal set of necessary $6j$ symbols, namely those
appearing in the different steps above,
Eqs.~(\ref{eq:CRApplicationLoopType1}-\ref{eq:CRApplicationLoopType1LastCR})
  and \Eqref{eq:CRApplicationLoopType3}.
Keeping in mind that the single lines above denote adjoint or
fundamental representations, we conclude that the $6j$s we are after
can be divided into the cases in \tabref{tab:6js}.

\begin{table}
\begin{tabular}{|l|l|l|l|}
  \hline
  
    \hspace*{0.7 cm}$n_g=0$ & \hspace*{2 cm}$n_g=1$ & \hspace*{0.7 cm}$n_g=2$ & \hspace*{2 cm}$n_g=3$\\\hline
  
  $\FPic[scale=0.92]{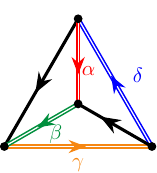}$
  & $\FPic[scale=0.92]{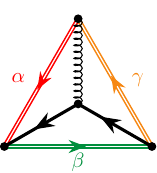}$
  $\FPic[scale=0.92]{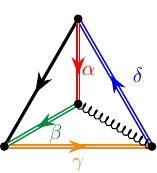}$
  &  $\FPic[scale=0.92]{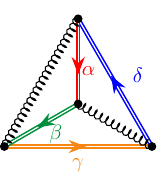}$
  &  $\FPic[scale=0.92]{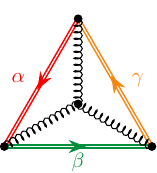}\
  \FPic[scale=0.92]{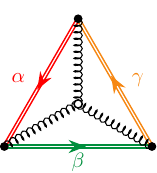}$
  \\\hline
  \hspace*{0.7 cm} case 0 & \hspace*{0.7 cm} case 1 \hspace*{0.5 cm} | \hspace*{0.5 cm} case 2 & \hspace*{0.7 cm} case 3 & \hspace*{1.9 cm} case 4
  \\\hline
\end{tabular}
\caption{The required set of $6j$ symbols for color decomposition into
  multiplet bases of the form in \eqref{eq:basis vector}.
  The last two $6j$s have the antisymmetric ($f$) and symmetric ($d$) triple-gluon vertices
  in the middle respectively.
  \label{tab:6js}}
\end{table}

We note that the $6j$s of type (0) in \tabref{tab:6js} are known
from Ref. \citenum{Alcock-Zeilinger:2022hrk}. In this article we address the computation of
the remaining $6j$s. Before taking on this task, we must, however,
be careful with how we define the vertices for the cases where we
have more than one vertex between the same set of representations,
which can happen for vertices with gluons, as discussed below
\Eqref{eq:vertex-correction}.

\section{Vertex construction}
\label{sec:vertices}

In \tabref{tab:6js} we sorted the $6j$ symbols that we are going to
study in this work according to the number of gluon lines and
according to the number of vertices with gluon lines that these $6j$
symbols contain. Before we can evaluate the $6j$ symbols, we have to
construct all vertices with at least one gluon line. When discussing
how many vertices with a given set of irreps there are, it
is useful to think of the general irrep labels (for which we use Greek
letters) as Young diagrams. In fact, a systematic labeling of $\SUN$
irreps applicable for \textit{arbitrary} $N$ should rather be in terms of pairs
of Young diagrams.\cite{King:1970,Sjodahl:2018cca} However, if we allow for Young
diagrams with columns with an $N$-dependent number of boxes, we can
replace each pair of Young diagrams by a single Young diagram 
\cite{Sjodahl:2018cca}. For instance, the
adjoint representation is then labeled by the Young diagram
$A=\begin{ytableau} *(black) \ & \ \\ *(black) \ \\ *(black)
\ \end{ytableau}$, where, here and in what follows, a black column
always represents a column with $N-1$ boxes. Hence, in the following,
we can always think of irrep labels as single Young diagrams with,
possibly, $N$-dependent column lengths.

We will normalize all our vertices such that all non-vanishing $3j$
symbols are equal to $1$, as already mentioned following
\Eqref{eq:scalar-product-and-3j}. Readers who prefer to work with
different normalizations are referred to \appref{app:restore-3j} for a
simple transformation rule.

For each instance of the irrep $\alpha$ in the complete reduction of
$\gamma \otimes A$ we have to construct a vertex
\begin{equation}
  \label{eq:Vertex-alphaO-adj-gammaI}
  \FPic{Vertex-alphaO-adj-gammaI--Va} \, ,
\end{equation}
where $a$ thus runs from 1 to the multiplicity of $\alpha$ in $\gamma\otimes A$. 
Essentially, we construct these vertices by splitting the gluon line into a
$q\bar{q}$-pair. More precisely, we consider all diagrams
\begin{equation}
  \label{eq:Vertex-alphaO-adj-gammaI--lambdaj-fund-fundST}
  \FPic{Vertex-alphaO-adj-gammaI--lambdaj-fund-fundSTAR} \, , 
\end{equation}
where $j$ enumerates admissible intermediate irreps according to a
scheme to be explained below, and construct the desired vertices as
linear combinations of these diagrams. In general, these vertices then
take the form
\begin{equation}
  \FPic{Vertex-alphaO-adj-gammaI--Va} 
  = \sum_j C^{\alpha\gamma}_{aj}
  \FPic{Vertex-alphaO-adj-gammaI--lambdaj-fund-fundSTAR}
\end{equation}
with coefficients $C^{\alpha\gamma}_{aj} \in \mathbb{C}$, which will
actually turn out to be real-valued functions of $N$. We distinguish
the two cases $\alpha\neq\gamma$ and $\alpha=\gamma$.

If $\alpha\neq\gamma$ then there is only a non-zero vertex
\eqref{eq:Vertex-alphaO-adj-gammaI} if $\alpha$ can be found in
the Clebsch-Gordan decomposition of $\gamma \otimes A$.
This means that we can obtain
$\alpha$ from
$\gamma$ by adding a box in one row and subsequently removing a box in
a different row (possibly after first adding a column of
length $N$)\cite{Keppeler:2012ih}. In this case there is a unique intermediate irrep
$\lambda_1$ in diagram
\eqref{eq:Vertex-alphaO-adj-gammaI--lambdaj-fund-fundST},
representing the intermediate step in this process after adding a box
but before removing the other box. Hence, for such $\alpha\neq\gamma$, we
find
\begin{equation}
  \label{eq:vertex_for_alpha_neq_gamma}
  \FPic{Vertex-alphaO-adj-gammaI--V1}
  =
  C^{\alpha\gamma}_{11} \FPic{Vertex-alphaO-adj-gammaI--lambda1-fund-fundSTAR}
  \, , 
\end{equation}
where the constant has to be chosen such that the normalization
condition for the corresponding $3j$ symbol,
\begin{equation}
  \scalebox{1.5}{\Bigg\langle}
  \FPic{Vertex-alphaO-adj-gammaI--V1}
  \ , \
  \FPic{Vertex-alphaO-adj-gammaI--V1}
  \scalebox{1.5}{\Bigg\rangle}
  = \FPic{3j-alpha-gamma-adj--V1-V1}
  \stackrel{!}{=} 1
  \, ,
\end{equation}
is fulfilled.
After a few steps, spelled out in \appref{App:props_of_vertex-corr-diagrams},
we get from \Eqref{eq:square_diagram_final_result}
\begin{equation}
  \label{eq:different rep norm}
  C^{\alpha\gamma}_{11}=\sqrt{d_{\lambda_1}(N^2-1)} \quad \text{for} \quad \alpha\neq\gamma \;.
\end{equation}

For $\alpha=\gamma$ there can be up to $N-1$ vertices of type
\eqref{eq:Vertex-alphaO-adj-gammaI}, cf.\ Appendix~B of
Ref.~\citenum{Keppeler:2012ih},
i.e., if the multiplicity of
$\alpha$ in the complete reduction of $\alpha\otimes A$
is $K$, we have to construct the vertices
\begin{equation}
  \label{eq:Vertex-alphaO-adj-alphaI--Va}
  \FPic{Vertex-alphaO-adj-alphaI--Va} \ , \quad a=1,\ldots,K\, .
\end{equation}
In this case there exist $K+1$ different admissible irreps $\lambda_j$
rendering the diagrams
\begin{equation}
  \label{eq:Vertex-alphaO-adj-alphaI--lambdaj-fund-fundSTAR}
  \FPic{Vertex-alphaO-adj-alphaI--lambdaj-fund-fundSTAR}
  \ , \quad j=1,\ldots,K+1 \, , 
\end{equation}
non-zero, as discussed in Appendix~\ref{App:props_of_vertex-corr-diagrams}, where
we also show that all vertices in \Eqref{eq:Vertex-alphaO-adj-alphaI--Va}
are linear combinations of theses diagrams, i.e.\
\begin{equation}
  \label{eq:vertices_for_alpha=gamma}
  \FPic{Vertex-alphaO-adj-alphaI--Va}
  = 
  \sum_{j} C^{\alpha\alpha}_{aj} 
  \FPic{Vertex-alphaO-adj-alphaI--lambdaj-fund-fundSTAR} \, .
\end{equation}
Hence, we can obtain a set of orthonormal
vertices~\eqref{eq:Vertex-alphaO-adj-alphaI--Va} by applying the
Gram-Schmidt algorithm to the set of
diagrams~\eqref{eq:Vertex-alphaO-adj-alphaI--lambdaj-fund-fundSTAR}
with admissible intermediate irreps $\lambda_j$.

In order to obtain a unique result when carrying out Gram-Schmidt we
have to decide how to sort the
diagrams~\eqref{eq:Vertex-alphaO-adj-alphaI--lambdaj-fund-fundSTAR}. To
this end, note that
an admissible $\lambda_j$ is obtained by adding an extra box to 
$\alpha$. We say that
\begin{equation}
  \FPic{Vertex-alphaO-adj-alphaI--lambdaj-fund-fundSTAR}
  \quad < \quad
  \FPic{Vertex-alphaO-adj-alphaI--lambdak-fund-fundSTAR}
\end{equation}
if in
$\lambda_k$ this extra box is added
further down compared to where it was added in
$\lambda_j$. We then sort the birdtrack
diagrams~\eqref{eq:Vertex-alphaO-adj-alphaI--lambdaj-fund-fundSTAR} in
increasing order. Hence, the first birdtrack diagram in our list is
always the diagram with intermediate irrep $\lambda_1$
which is obtained by adding a box to the first row of
$\alpha$. In Appendix~\ref{App:props_of_vertex-corr-diagrams} we show
that the last diagram in this list is always a linear combinations of
the first $K$ diagrams and can thus be omitted. The sum in 
\Eqref{eq:vertices_for_alpha=gamma} hence runs from $1$ to $K$.

In order to carry out Gram-Schmidt we only need to know the scalar
products between all
diagrams~\eqref{eq:Vertex-alphaO-adj-alphaI--lambdaj-fund-fundSTAR},
which are calculated in App.~\ref{App:props_of_vertex-corr-diagrams},
Eq.~\eqref{eq:scalar_product_is_square_diagram}--\eqref{eq:square_diagram_final_result}.
We denote this
\begin{equation}
\label{eq:scalar_product_of_two_vertex_corrections}
  s_{jk} =
  \scalebox{1.5}{\Bigg\langle}
  \FPic{Vertex-alphaO-adj-alphaI--lambdaj-fund-fundSTAR}
  \ , \
  \FPic{Vertex-alphaO-adj-alphaI--lambdak-fund-fundSTAR}
  \scalebox{1.5}{\Bigg\rangle}
  = \frac{1}{N^2-1} \left(
    \frac{\delta_{jk}}{d_{\lambda_j}} - \frac{1}{N d_\alpha} \right)\;.   
\end{equation}
The scalar products $s_{jk}$ also depend on the irrep
$\alpha$ but we do not display this dependence in our notation since
in the following $s_{jk}$ for different irreps $\alpha$ never appear
alongside each other in our equations.

We explicitly state the formulae
for the first two vertices, which are the only vertices in the
physically relevant case $N=3$,
\begin{align}
  \label{eq:Vertex-alphaO-adj-alphaI--V1}
  \FPic{Vertex-alphaO-adj-alphaI--V1}
  & = \frac{1}{\sqrt{s_{11}}}
  \FPic{Vertex-alphaO-adj-alphaI--lambda1-fund-fundSTAR}
  \\
  \label{eq:Vertex-alphaO-adj-alphaI--V2}
  \FPic{Vertex-alphaO-adj-alphaI--V2}
  & = \sqrt{\frac{s_{11}}{s_{11}s_{22}-{s_{12}}^2}} \, 
  \left( \FPic{Vertex-alphaO-adj-alphaI--lambda2-fund-fundSTAR}
  - \frac{s_{12}}{s_{11}}
  \FPic{Vertex-alphaO-adj-alphaI--lambda1-fund-fundSTAR} \right)
\end{align}
Further vertices, which only exist for $N>3$, are calculated by
straightforwardly continuing Gram-Schmidt.
In \appref{app:examples} we illustrate the explicit vertex
construction with a few examples.

\section{Formulae for gluon~\texorpdfstring{$6j$}{6j}~symbols}
\label{sec:closed-form-expressions}

We will now describe how the calculation of the different classes of
$6j$ symbols proceeds. Our main tool will be the repeated insertion of
vertex corrections, and the Fierz identity, \Eqref{eq:Fierz}, to decompose gluon
lines. We will go through the basic idea and steps for the simpler
cases here, but defer the details of longer calculations to
\appref{App:steps} for the sake of readability.

\subsection*{Case 1: $6j$s with a quark-gluon vertex}

We here consider the $6j$ symbol which contains a $q\bar{q} g$ vertex (in its
center). We proceed to calculate this by expanding the gluon vertex into a
vertex correction

\begin{equation}
  \label{eq:case1a}
  \FPic{6j-alpha-beta-gamma--fundSTAR-fund-adj}
  \xlongequal{\eqref{eq:vertex_for_alpha_neq_gamma}}
  \sqrt{d_{\lambda_1}(N^2-1)} \
  \FPic{Square-alpha-beta-gamma-lambda1--fundSTAR-fund-fundSTAR-fund-adj}
  \xlongequal{\eqref{eq:square_diagram_final_result}}
  \frac{\delta_{\beta\lambda_1}}{\sqrt{d_\beta(N^2-1)}},
\end{equation}
where we have assumed $\alpha\neq\gamma$ and used
\Eqref{eq:square_diagram_final_result} from
\appref{App:props_of_vertex-corr-diagrams}, which builds on the Fierz
identity, \Eqref{eq:Fierz}.

In the case $\alpha=\gamma$, again using
\Eqref{eq:square_diagram_final_result}, for the first two
vertices $a=1,2$ we obtain 
\begin{align}
  \label{eq:case1b}
  \FPic{6j-alpha-lambdak-alpha--fundSTAR-fund-adj--V1}
  &\xlongequal{\eqref{eq:Vertex-alphaO-adj-alphaI--V1}}
  \frac{s_{1k}}{\sqrt{s_{11}}} \qquad \text{and}
  \\
  \FPic{6j-alpha-lambdak-alpha--fundSTAR-fund-adj--V2}
  &\xlongequal{\eqref{eq:Vertex-alphaO-adj-alphaI--V2}}
  \sqrt{\frac{s_{11}}{s_{11}s_{22}-{s_{12}}^2}} 
  \left( s_{2k} - \frac{s_{12} }{s_{11}}s_{1k}\right) \, .
\end{align}
Note that the last expression vanishes for $k=1$. For $N>3$,
there might, as described, be more vertices which then
are treated similarly.

\subsection*{Case 2: $6j$s with a gluon line opposing a quark line}

In this case we have one quark and one gluon line attaching to
different vertices. Rewriting the gluon vertices in terms of
vertex corrections and invoking the Fierz identity, we then find,
after a few steps spelled out in \appref{App:steps} 
\begin{equation}
  \label{eq:case 2}
\begin{split}
  &\FPic{6j-fund-gamma-delta--betaSTAR-adj-alpha--Va-Vb_smaller}
   = \sum_{j=1}^a \sum_{k=1}^b
     \frac{C^{\beta\alpha}_{aj} C^{\delta\gamma}_{bk}}{N^2-1} \left(    
           \FPic{6j-fund-muk-delta--lambdajSTAR-fund-alpha}
           \FPic{6j-fund-gamma-muk--betaSTAR-fundSTAR-lambdaj}
     - \frac{\delta_{\alpha\beta} \, \delta_{\gamma\delta}}
            {N d_\alpha \, d_\gamma} 
       \right)\;,
\end{split}
\end{equation}
where the $6j$ symbols with two quark lines are given in closed form 
in Ref.~\citenum{Alcock-Zeilinger:2022hrk}.

\subsection*{Case 3: $6j$s with two opposing gluon lines}

Case 3 can be addressed with a similar strategy as the other cases. Our
result, for which we demonstrate all intermediate steps in
\appref{App:steps}, reads
\begin{equation}
  \label{eq:case 3}
\begin{split}
  &\FPic{6j-adj-gamma-delta--betaSTAR-adj-alpha--Va-Vb-Vc-Vd_smaller}
  =  \sum_{j=1}^a \sum_{k=1}^b
     \frac{C^{\beta\alpha}_{aj} C^{\delta\gamma}_{bk}}{N^2-1} \left(    
           \FPic{6j-fund-alphaSTAR-deltaSTAR--lambda-adj-mukSTAR--Vc-Vd}
           \FPic{6j-fund-gammaSTAR-betaSTAR--muk-adj-lambdajSTAR--Vc-Vd}
     - \frac{\delta_{\alpha\beta} \, \delta_{\gamma\delta}}
            {N d_\alpha \, d_\gamma}
       \right)\;.
     \end{split}
\end{equation}

\subsection*{Case 4: $6j$s with three-gluon vertices}

The class of the $6j$ symbols with three gluons consists of two three-gluon
vertices, which typically is taken to be proportional to $if^{abc}$ and
$d^{abc}$. We will illustrate the case of $if^{abc}$ first.
 In particular, we use the definition of the $if^{abc}$-vertex in terms
of traces, and then insert vertex corrections,
\begin{equation}
\begin{split}
  \FPic{6j-alpha-beta-gamma--adj-adj-adj--Vf-Va-Vb-Vc_smaller}
  &= \frac{N^2-1}{\sqrt{2N}} \scalebox{2.2}{\Bigg(}
  \FPic{6j-alpha-beta-gamma--adj-adj-adj--loop-clockwise--Va-Vb-Vc_smaller}
  -
  \FPic{6j-alpha-beta-gamma--adj-adj-adj--loop-counterclockwise--Va-Vb-Vc_smaller}  
  \scalebox{2.2}{\Bigg)}
\end{split}
\end{equation}
Using the Fierz identity, \Eqref{eq:Fierz}, we can remove all internal
gluon lines and the results are expressed in terms of a number of
different diagrams which reduce to $3j$ symbols, dimensions and traces
over quark lines, see Appendix~\ref{App:steps}. A single non-trivial
diagram remains, which can be expressed as
\begin{equation}
\begin{split}
  \FPic{6j-alpha-beta-gamma--insert-lambdaj-muk-nul--crossing}
  &=  \sum_{\sigma} d_\sigma \, 
  \FPic{6j-alpha-lambdaj-fund--fundSTAR-beta-sigmaSTAR}
  \FPic{6j-fund-muk-gamma--betaSTAR-fund-sigma}
  \FPic{6j-fund-nul-alpha--gammaSTAR-fund-sigma}\;,
\end{split}
\end{equation}
where a minus sign next to a vertex indicates that the lines are
connected to this vertex in opposite order, i.e.\
\begin{equation}
  \label{eq:signed-vertex}
  \FPic{GenRep-VertexSTAR-gammaO-betaO-alphaO}
  \; = \; 
  \FPic{GenRep-Vertex-gammaO-betaOSTAR-alphaOSTAR}
  \ ,
\end{equation}
see also appendix~C of Ref.~\citenum{Alcock-Zeilinger:2022hrk}. For
the above vertices with quarks this only makes a difference for the
antisymmetic vertex of $q\otimes q$.\cite{Alcock-Zeilinger:2022hrk}
Our final result for the $f$-vertex is
\begin{multline}
  \label{eq:f}
  \FPic{6j-alpha-beta-gamma--adj-adj-adj--Vf-Va-Vb-Vc_smaller}
  = \frac{1}{(N^2-1)^2}\frac{1}{\sqrt{2N}} \sum_{j=1}^a \sum_{k=1}^b \sum_{\ell=1}^c
  C^{\beta\alpha}_{aj} C^{\gamma\beta}_{bk} C^{\alpha\gamma}_{c\ell}\\
  \scalebox{2.2}{\Bigg(}
  \sum_{\sigma} d_\sigma \, 
  \FPic{6j-alpha-lambdaj-fund--fundSTAR-beta-sigmaSTAR}
  \FPic{6j-fund-muk-gamma--betaSTAR-fund-sigma}
  \FPic{6j-fund-nul-alpha--gammaSTAR-fund-sigma}
  \ - \
  \frac{\delta_{\lambda_j\mu_k}\,\delta_{\lambda_j\nu_\ell}}{d_{\lambda_j}^2}
  \scalebox{2.2}{\Bigg)} \ ,
\end{multline}
while we obtain for the $d$-vertex,
\begin{multline}
  \label{eq:d}
  \FPic{6j-alpha-beta-gamma--adj-adj-adj--Vd-Va-Vb-Vc_smaller}
  = \frac{1}{(N^2-1)^2}\sqrt{\frac{N}{2(N^2-4)}} \sum_{j=1}^a \sum_{k=1}^b \sum_{\ell=1}^c
  C^{\beta\alpha}_{aj} C^{\gamma\beta}_{bk} C^{\alpha\gamma}_{c\ell}\\
  \scalebox{2.2}{\Bigg(}
\sum_{\sigma} d_\sigma \, 
  \FPic{6j-alpha-lambdaj-fund--fundSTAR-beta-sigmaSTAR}
  \FPic{6j-fund-muk-gamma--betaSTAR-fund-sigma}
  \FPic{6j-fund-nul-alpha--gammaSTAR-fund-sigma}
  \ + \
    \frac{\delta_{\lambda_j\mu_k}\,\delta_{\lambda_j\nu_\ell}}{d_{\lambda_j}^2}
  \\
 + \frac{4}{N^2} \frac{\delta_{\alpha\beta}\,\delta_{\alpha\gamma}}{d_\alpha^2}
  - \frac{2}{N}\left(
    \frac{\delta_{\alpha\gamma}\,\delta_{\lambda_j\mu_k}}
    {d_\alpha \, d_{\lambda_j}}
    + \frac{\delta_{\alpha\beta}\,\delta_{\mu_k\nu_\ell}}
    {d_\alpha \, d_{\mu_k}}
    + \frac{\delta_{\beta\gamma}\,\delta_{\lambda_j\nu_\ell}}
    {d_\beta \, d_{\lambda_j}}\right)
  \scalebox{2.2}{\Bigg)} \ .
\end{multline}
\section{Conclusions and outlook}
\label{sec:conclusions}

In the present paper we have shown how to calculate a
set of Wigner $6j$ coefficients with adjoint representations.
Together with a set of previously derived $6j$s\cite{Alcock-Zeilinger:2022hrk},
this set constitutes a complete set of $6j$s required to decompose any
color structure, to any order into orthogonal multiplet
bases, cf. \Eqref{eq:basis vector}.

This opens up for the usage of orthogonal representation theory based
color bases also for processes with high multiplicities, including the
analysis of evolution equations in color space
\cite{Platzer:2022jny}.

We note, however, that the present work does not close
the research area of representation theory based treatment
of color structure. In particular, more general $6j$ symbols
are required for fully general multiplet bases (with vertices
between general representations). We believe that this
can be addressed with similar methods.

\section*{Acknowledgments}
We are thankful to Judith Alcock-Zeilinger for useful
discussions.
We also thank the Erwin Schrödinger International
Institute for Mathematics and Physics (ESI) in Vienna, for hospitality,
discussions, and support, both via the Research in Teams programme
“Amplitude Level Evolution II: Cracking down on color bases” (RIT0521),
where this work was initiated, and via the ESI-QFT 2023 workshop,
where the scientific part was concluded.
MS acknowledges support by the Swedish Research Council (contract
number 2016-05996, as well as the European Union’s Horizon 2020
research and innovation programme (grant agreement No 668679).

\appendix

\section{Properties of vertex correction diagrams}
\label{App:props_of_vertex-corr-diagrams}

We discuss some properties of the vertex correction
diagrams in \Eqref{eq:Vertex-alphaO-adj-gammaI--lambdaj-fund-fundST},
which we use for the construction of vertices with at least one gluon.
Let $K$ be the multiplicity of $\alpha$ in the complete reduction
of $\alpha\otimes A$, giving the number of
vertices~\eqref{eq:Vertex-alphaO-adj-alphaI--Va} to be constructed. In
contrast, the number of intermediate irreps $\lambda_j$ in
\Eqref{eq:Vertex-alphaO-adj-alphaI--lambdaj-fund-fundSTAR} is given
by the multiplicity of $\alpha$ in the complete reduction of
$\alpha\otimes\ydiagram{1}\otimes\overline{\ydiagram{1}}$. The latter
number is one higher than the former since due to
$\ydiagram{1}\otimes\overline{\ydiagram{1}}=A\otimes\bullet$ (where
$\bullet$ denotes the trivial representation) we have
$\alpha\otimes\ydiagram{1}\otimes\overline{\ydiagram{1}}
=(\alpha\otimes A) \oplus \alpha$. Moreover, $K+1$ is also the number
of terms in the complete reduction of $\alpha \otimes \ydiagram{1}$,
i.e.\ the number of ways in which we can add a box to the Young
diagram $\alpha$.

First we show that the
diagrams in \Eqref{eq:Vertex-alphaO-adj-alphaI--lambdaj-fund-fundSTAR}
are linearly dependent. To this end, consider the complete reduction of
$\alpha \otimes \ydiagram{1}$,
\begin{equation}
  \FPic{Misc-alpha-tensor-q}
  = \sum_{j=1}^{K+1} d_{\lambda_j} \FPic{Misc-project-alpha-q-to-lambdaj} \ ,
\end{equation}
multiply with a quark-gluon vertex (Lie algebra generators), and
contract the quark and antiquark lines, yielding
\begin{equation}
  \label{eq:linear_dependence_of_vertex_corrections}
  \underbrace{\FPic{Misc-alpha-tadpole}}_{=0}
  = \sum_{j=1}^{K+1} d_{\lambda_j} 
  \FPic{Vertex-alphaO-adj-alphaI--lambdaj-fund-fundSTAR} \, .
\end{equation}
The l.h.s.\ vanishes since the $\SUN$ generators are traceless, i.e.\ we
have found a non-trivial vanishing linear combination of the
diagrams in \Eqref{eq:Vertex-alphaO-adj-alphaI--lambdaj-fund-fundSTAR}.

Next we show
that all vertices in \Eqref{eq:Vertex-alphaO-adj-alphaI--Va} are linear
combinations of the
diagrams in \Eqref{eq:Vertex-alphaO-adj-alphaI--lambdaj-fund-fundSTAR}. To
this end, consider a gluon exchange between $\alpha$ and a quark line,
and insert two completeness relations:
\begin{equation}
\label{eq:Misc-alpha-q-gluon_exchange}
  \FPic{Misc-alpha-q-gluon_exchange}
  = \sum_{j,k=1}^{K+1} d_{\lambda_j} \, d_{\lambda_k}
  \FPic{Misc-alpha-q-gluon_exchange-inserted_completness}
  \ . 
\end{equation}
Due to Schur's Lemma the middle segment can only be non-zero if
$\lambda_j$ and $\lambda_k$ are equivalent. If two irreps in the
complete reduction of $\alpha\otimes\ydiagram{1}$ are equivalent then
they are the same, i.e.
\begin{equation}
  \FPic{Misc-alpha-q-gluon_exchange-middle_section}
  = \widetilde{C}_{aj} \, \delta_{jk} \FPic{Misc-lambdaj}
\end{equation}
with some constant $\widetilde{C}_{aj}$. Substituting into
\Eqref{eq:Misc-alpha-q-gluon_exchange}, multiplying with a
quark-gluon vertex, and contracting the quark and antiquark line we
find
\begin{equation}
  \FPic{Misc-alpha-gluon_w_quark_loop}
  = \sum_{j=1}^{K+1} \widetilde{C}_{aj} \, d_{\lambda_j}^2 \, 
  \FPic{Vertex-alphaO-adj-alphaI--lambdaj-fund-fundSTAR} \, .
\end{equation}
The quark loop on the l.h.s.\ can be traded for a factor of
$(N^2-1)^{-1}$ (recall that we set all $3j$ symbols equal to $1$), and by
defining $C^{\alpha\alpha}_{aj} = \widetilde{C}_{aj} \, d_{\lambda_j}^2 \, (N^2-1)$
we obtain
\Eqref{eq:vertices_for_alpha=gamma}, as claimed in
\secref{sec:vertices}.

Now we can even take advantage of the linear
dependence~\eqref{eq:linear_dependence_of_vertex_corrections} of the
vertex correction
diagrams~\eqref{eq:Vertex-alphaO-adj-alphaI--lambdaj-fund-fundSTAR}. Equation~\eqref{eq:linear_dependence_of_vertex_corrections}
tells us that any one of the $K+1$
diagrams~\eqref{eq:Vertex-alphaO-adj-alphaI--lambdaj-fund-fundSTAR}
can be expressed as a linear combination of the other $K$ diagrams,
since none of the coefficients in
Eq.~\eqref{eq:linear_dependence_of_vertex_corrections} vanishes. In
Sec.~\ref{sec:vertices} we order the diagrams in a unique way and
determine the orthonormal
vertices in \Eqref{eq:Vertex-alphaO-adj-alphaI--Va} by means of the
Gram-Schmidt algorithm. Since the last vertex correction diagram is
guaranteed to be a linear combination of the first $K$ diagrams, we
can always terminate Gram-Schmidt \textit{before} using the last
diagram, i.e\ the vertices~\eqref{eq:Vertex-alphaO-adj-alphaI--Va} are
actually linear combinations of the first $K$ vertex correction
diagrams~\eqref{eq:Vertex-alphaO-adj-alphaI--lambdaj-fund-fundSTAR}.
 
Finally, we explicitly determine the scalar products between all
vertex correction
diagrams~\eqref{eq:Vertex-alphaO-adj-gammaI--lambdaj-fund-fundST}. The
result is the main ingredient for the Gram-Schmidt process in
Sec.~\ref{sec:vertices}. Consider
\begin{equation}
  \label{eq:scalar_product_is_square_diagram}
  \scalebox{1.5}{\Bigg\langle}
  \FPic{Vertex-alphaO-adj-gammaI--beta-fund-fundSTAR}
  \ , \
  \FPic{Vertex-alphaO-adj-gammaI--delta-fund-fundSTAR} 
  \scalebox{1.5}{\Bigg\rangle}
  = 
  \FPic{Square-alpha-beta-gamma-delta--fundSTAR-fund-fundSTAR-fund-adj}
  \, .
\end{equation}
The square diagram can be evaluated by invoking
the Fierz identity (or adjoint representation projector, also
equivalent to the completeness relation for $q\otimes \bar{q}$), which
with our unit $3j$ symbols takes the form
\begin{equation}
\label{eq:Fierz}
\raisebox{-0.8 cm}{\includegraphics[scale=0.5]{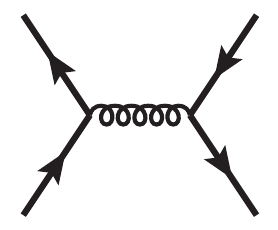}}
=
 \frac{1}{N^2-1}\left(
\raisebox{-0.8 cm}{\includegraphics[scale=0.5]{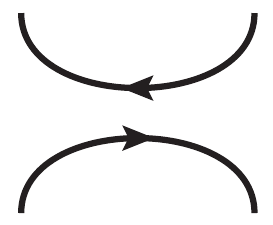}}
 - \frac{1}{N}
\raisebox{-0.8 cm}{\includegraphics[scale=0.5]{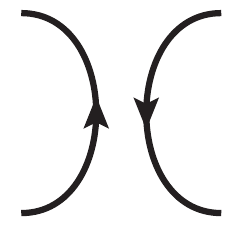}}\right)\;.
\end{equation}
Inserting this gives for the scalar product
\begin{equation}\label{eq:square_diagram_final_result}
\begin{split}
  \FPic{Square-alpha-beta-gamma-delta--fundSTAR-fund-fundSTAR-fund-adj}
  \ &= \
  \frac{1}{N^2-1}
  \left(
  \ \FPic{Square-alpha-beta-gamma-delta--vert} \
  - \frac{1}{N}
  \ \FPic{Square-alpha-beta-gamma-delta--horiz} \
  \right)
  \\[1ex]
  \ &= \
  \frac{1}{N^2-1} \left(
  \frac{\delta_{\beta\delta}}{d_\beta}
  - \frac{\delta_{\alpha\gamma}}{Nd_\alpha} 
  \right) \, .
  \end{split}
\end{equation}

\section{Examples of vertex construction}
\label{app:examples}

We illustrate how to construct vertices of type
\eqref{eq:Vertex-alphaO-adj-gammaI} using methods and results from
\secref{sec:vertices}.

First consider an example with $\alpha\neq\gamma$.  For
$\alpha=\ydiagram{1,1}$ and $\gamma=\ydiagram{2}$ the unique
intermediate irrep is $\lambda_1=\ydiagram{2,1}$. Then, using
\Eqref{eq:different rep norm} for normalization, the unique vertex with irreps
$\ydiagram{1,1}$, $\ydiagram{2}$ and one gluon reads
\begin{equation}
  \FPic{Vertex-Y11-adj-Y2--V1} =
  (N^2-1) \sqrt{\frac{N}{3}} \ \FPic{Vertex-Y11-adj-Y2--Y21-fund-fundSTAR}
  \, .
\end{equation}

The Young diagram with the smallest number of boxes for which there is more
than one vertex is 
$\alpha=\gamma=\ydiagram{2,1}$, i.e., the octet for $N=3$ (note that
this is \textit{not} the adjoint representation for $N\ne3$). The admissible
intermediate irreps are $\lambda_1=\ydiagram{3,1}\,$ and
$\lambda_2=\ydiagram{2,2}\;$. Using
\Eqref{eq:Vertex-alphaO-adj-alphaI--V2}, the orthonormal vertices
become
\begin{align}
  \label{eq:Y21-in-out-first-vertex}
  \FPic{Vertex-Y21-adj-Y21--V1}
  & = N (N^2-1) \sqrt{\frac{N+2}{5N-6}} \
  \FPic{Vertex-Y21-adj-Y21--Y31-fund-fundSTAR} \qquad \text{and}
  \\
  \label{eq:Y21-in-out-second-vertex}
  \FPic{Vertex-Y21-adj-Y21--V2}
  & = \frac{N(N^2-1)}{6} \sqrt{\frac{5N-6}{N-2}} \, 
  \left( \FPic{Vertex-Y21-adj-Y21--Y22-fund-fundSTAR}
  + 3\frac{N+2}{5N-6}\
  \FPic{Vertex-Y21-adj-Y21--Y31-fund-fundSTAR} \right) \, .
\end{align}

For $\alpha=\gamma=A=\begin{ytableau} *(black) \ & \ \\ *(black)
\ \\ *(black) \ \end{ytableau}$ (recall that the black column stands
for a column with $N-1$ boxes) we obtain three-gluon vertices for general $N$. The
admissible intermediate irreps are then $\lambda_1=\begin{ytableau}
*(black) \ & \ & \ \\ *(black) \ \\ *(black) \ \end{ytableau}$ and
$\lambda_2=\begin{ytableau} *(black) \ & \ \\ *(black) \ &
\ \\ *(black) \ \end{ytableau}\;$.  Using
\Eqref{eq:Vertex-alphaO-adj-alphaI--V2}, our orthonormal vertices read
\begin{align}
  \label{eq:adj-in-out-first-vertex}
  \FPic{Vertex-adj-adj-adj--V1}
  & = (N^2-1) \sqrt{N+2} \
  \FPic{Vertex-adj-adj-adj--Yc2-fund-fundSTAR} \qquad \text{and}
  \\
  \label{eq:adj-in-out-second-vertex}
  \FPic{Vertex-adj-adj-adj--V2}
  & = \frac{N(N^2-1)}{2} \sqrt{N-2} \, 
  \left( \FPic{Vertex-adj-adj-adj--Yc11-fund-fundSTAR}
  + \frac{N+2}{N}\
  \FPic{Vertex-adj-adj-adj--Yc2-fund-fundSTAR} \right) \, .
\end{align}
Notice that for $N=3$
Eqs.~\eqref{eq:Y21-in-out-first-vertex}/\eqref{eq:Y21-in-out-second-vertex}
and
Eqs.\eqref{eq:adj-in-out-first-vertex}/\eqref{eq:adj-in-out-second-vertex}
coincide. Instead of the latter vertices, one will likely want to use
the much more common antisymmetric $f$ and symmetric $d$ vertices, to
which our vertices are related by a unitary transformation, which we
explicitly state below.
Like all other vertices in this article we normalize $f$ and $d$
such that the corresponding $3j$ symbols are equal to one, i.e.
\begin{align}
  \FPic{Vertex-adj-adj-adj--Vf}
  \ &= \
  \frac{N^2-1}{\sqrt{2N}} \left( \FPic{Vertex-adj-adj-adj--loop-clockwise}
  - \FPic{Vertex-adj-adj-adj--loop-counterclockwise} \right)
  \qquad \text{and}
  \\
  \FPic{Vertex-adj-adj-adj--Vd}
  \ &= \
  (N^2-1) \sqrt{\frac{N}{2(N^2-4)}} \left(
  \FPic{Vertex-adj-adj-adj--loop-clockwise}
  + \FPic{Vertex-adj-adj-adj--loop-counterclockwise} \right) \, ,
\end{align}
see \appref{app:restore-3j} for how to easily transform results to
other normalizations. The vertices \eqref{eq:adj-in-out-first-vertex}
and \eqref{eq:adj-in-out-second-vertex} are related to $f$ and $d$ by
a unitary transformation,
\begin{align}
  \FPic{Vertex-adj-adj-adj--V1}
  \ &= \
  -\sqrt{\frac{N+2}{2N}} \FPic{Vertex-adj-adj-adj--Vf}
  + \sqrt{\frac{N-2}{2N}} \FPic{Vertex-adj-adj-adj--Vd} \, ,
  \\
  \FPic{Vertex-adj-adj-adj--V2}
  \ &= \
  - \sqrt{\frac{N-2}{2N}} \FPic{Vertex-adj-adj-adj--Vf}
  - \sqrt{\frac{N+2}{2N}} \FPic{Vertex-adj-adj-adj--Vd} \, , 
\end{align}
and vice versa,  
\begin{align}
  \FPic{Vertex-adj-adj-adj--Vf}
  \ &= \
  - \sqrt{\frac{N+2}{2N}} \FPic{Vertex-adj-adj-adj--V1}
  - \sqrt{\frac{N-2}{2N}} \FPic{Vertex-adj-adj-adj--V2} \, ,
  \\
  \FPic{Vertex-adj-adj-adj--Vd}
  \ &= \
  \hphantom{-} \sqrt{\frac{N-2}{2N}} \FPic{Vertex-adj-adj-adj--V1}
  - \sqrt{\frac{N+2}{2N}} \FPic{Vertex-adj-adj-adj--V2} \, ,
\end{align}
facilitating easy conversion.

The coefficients of this unitary transformation are determined by
scalar products between the two sets of vertices, and these scalar
products can be evaluated by calculations similar to
Eqs.~\eqref{eq:scalar_product_is_square_diagram}--\eqref{eq:square_diagram_final_result}.

\newpage
\section{Details of $6j$ derivations}
\label{App:steps}

We here give, in full detail, the intermediate steps for the
derivation in  \secref{sec:closed-form-expressions}.

\subsection*{Derivation for case 2}

We here derived the form of the $6j$ coefficients in \Eqref{eq:case 2}. 
In essence the vertices involving gluons are expressed in terms
of vertex corrections, after which the Fierz identity, \Eqref{eq:Fierz},
is applied, and vertex corrections are removed using
\Eqref{eq:vertex-correction}
\begin{equation}
\begin{split}
  &\FPic{6j-fund-gamma-delta--betaSTAR-adj-alpha--Va-Vb}
  = \sum_{j=1}^a \sum_{k=1}^b C^{\beta\alpha}_{aj} C^{\delta\gamma}_{bk}
     \FPic{6j-fund-gamma-delta--betaSTAR-adj-alpha--lambdaj-muk}
  \\
  &\quad = \sum_{j=1}^a \sum_{k=1}^b
     \frac{C^{\beta\alpha}_{aj} C^{\delta\gamma}_{bk}}{N^2-1} \left( 
     \FPic{6j-fund-gamma-delta--betaSTAR-adj-alpha--lambdaj-muk-connect}
     - \frac{1}{N}
     \FPic{6j-fund-gamma-delta--betaSTAR-adj-alpha--lambdaj-muk-loops}
     \right) 
  \\
  &\quad = \sum_{j=1}^a \sum_{k=1}^b
     \frac{C^{\beta\alpha}_{aj} C^{\delta\gamma}_{bk}}{N^2-1} \left(    
           \FPic{6j-fund-muk-delta--lambdajSTAR-fund-alpha}
           \FPic{6j-fund-gamma-muk--betaSTAR-fundSTAR-lambdaj}
     - \frac{\delta_{\alpha\beta} \, \delta_{\gamma\delta}}
            {N d_\alpha \, d_\gamma} 
       \right)\;.
\end{split}
\end{equation}

\newpage
\subsection*{Derivation for case 3}

The steps in the derivation of \Eqref{eq:case 3} progress
similarly to those in the derivation of \Eqref{eq:case 2},
\begin{equation}
\begin{split}
  &\FPic{6j-adj-gamma-delta--betaSTAR-adj-alpha--Va-Vb-Vc-Vd}
  = \sum_{j=1}^a \sum_{k=1}^b C^{\beta\alpha}_{aj} C^{\delta\gamma}_{bk}
     \FPic{6j-adj-gamma-delta--betaSTAR-adj-alpha--lambdaj-muk--Vc-Vd}
  \\
  &\quad = \sum_{j=1}^a \sum_{k=1}^b
     \frac{C^{\beta\alpha}_{aj} C^{\delta\gamma}_{bk}}{N^2-1} \left( 
     \FPic{6j-adj-gamma-delta--betaSTAR-adj-alpha--lambdaj-muk-connect--Vc-Vd}
     - \frac{1}{N}
     \FPic{6j-adj-gamma-delta--betaSTAR-adj-alpha--lambdaj-muk-loops--Vc-Vd}
     \right)
  \\
  &\quad = \sum_{j=1}^a \sum_{k=1}^b
     \frac{C^{\beta\alpha}_{aj} C^{\delta\gamma}_{bk}}{N^2-1} \left(    
           \FPic{6j-adj-muk-delta--lambdajSTAR-fund-alpha--Vc-Vd}
           \FPic{6j-adj-gamma-muk--betaSTAR-fundSTAR-lambdaj--Vc-Vd}
     - \frac{\delta_{\alpha\beta} \, \delta_{\gamma\delta}}
            {N d_\alpha \, d_\gamma}
       \right)
  \\
  &\quad = \sum_{j=1}^a \sum_{k=1}^b
     \frac{C^{\beta\alpha}_{aj} C^{\delta\gamma}_{bk}}{N^2-1} \left(    
           \FPic{6j-fund-alphaSTAR-deltaSTAR--lambda-adj-mukSTAR--Vc-Vd}
           \FPic{6j-fund-gammaSTAR-betaSTAR--muk-adj-lambdajSTAR--Vc-Vd}
     - \frac{\delta_{\alpha\beta} \, \delta_{\gamma\delta}}
            {N d_\alpha \, d_\gamma}
       \right)\;.
     \end{split}
\end{equation}
We remark that the result looks very similar to the result for case 2,
but that it is now expressed in terms of the $6j$s from case 2.

\newpage
\subsection*{Derivation for case 4}

Again the gluon vertices are expressed in terms of vertex corrections with
quarks, both in the triple-gluon vertices and in the vertices with the
general representations. This gives for the antisymmetric ($f$)
triple-gluon vertex
\begin{eqnarray}
    \label{eq:3g decomp}
  \FPic{6j-alpha-beta-gamma--adj-adj-adj--Vf-Va-Vb-Vc_smaller}
  &=& \frac{N^2-1}{\sqrt{2N}} \scalebox{2.2}{\Bigg(}
  \FPic{6j-alpha-beta-gamma--adj-adj-adj--loop-clockwise--Va-Vb-Vc_smaller}
  -
  \FPic{6j-alpha-beta-gamma--adj-adj-adj--loop-counterclockwise--Va-Vb-Vc_smaller}
  \scalebox{2.2}{\Bigg)}
  \\
  &=&  \frac{N^2-1}{\sqrt{2N}} \sum_{j=1}^a \sum_{k=1}^b \sum_{\ell=1}^c
  C^{\beta\alpha}_{aj} C^{\gamma\beta}_{bk} C^{\alpha\gamma}_{c\ell}
  \scalebox{2.2}{\Bigg(}
  \FPic{6j-alpha-beta-gamma--adj-adj-adj--loop-clockwise--insert-lambdaj-muk-nul}
  \ - \
  \FPic{6j-alpha-beta-gamma--adj-adj-adj--loop-counterclockwise--insert-lambdaj-muk-nul}
  \scalebox{2.2}{\Bigg)}\; \nonumber
\end{eqnarray}
and the symmetric ($d$) vertex differ only by the sign of the second term.

The second term above is calculated using the Fierz
identity~\eqref{eq:Fierz},
\begin{equation}
  \label{eq:3g first}
\begin{split}
  \FPic{6j-alpha-beta-gamma--adj-adj-adj--loop-counterclockwise--insert-lambdaj-muk-nul}
  &= \frac{1}{(N^2-1)^3} \scalebox{2.2}{\Bigg[}
  \FPic{6j-alpha-beta-gamma--insert-lambdaj-muk-nul--loops-alpha-beta-gamma}
  \\
  & \quad\, 
  - \frac{1}{N} \scalebox{2.2}{\Bigg(}
  \FPic{6j-alpha-beta-gamma--insert-lambdaj-muk-nul--loops-beta-nul}
  + \FPic{6j-alpha-beta-gamma--insert-lambdaj-muk-nul--loops-gamma-lambdaj}
  + \FPic{6j-alpha-beta-gamma--insert-lambdaj-muk-nul--loops-alpha-muk}
  \scalebox{2.2}{\Bigg)}
  \\
  & \quad\,
  + \frac{3}{N^2} \
  \FPic{6j-alpha-beta-gamma--insert-lambdaj-muk-nul--loops-lambdaj-muk-nul}
  - \frac{1}{N^3} \
  \FPic{6j-alpha-beta-gamma--insert-lambdaj-muk-nul--loops-lambdaj-muk-nul--counterclockwise}
  \scalebox{2.2}{\Bigg]}\;,
\end{split}
\end{equation}
where the closed quark loop in the last diagram simply yields a factor of $N$,
and the others are easy to evaluate using the self energy relation,
\Eqref{eq:self-energy},
for example
\begin{equation}
  \FPic{6j-alpha-beta-gamma--insert-lambdaj-muk-nul--loops-alpha-beta-gamma}
  = \frac{\delta_{\lambda_j\mu_k}\,\delta_{\lambda_j\nu_\ell}}{d_{\lambda_j}^2}\;.
\end{equation}

By identical steps, the first term in \Eqref{eq:3g decomp} gives
\begin{equation}
\begin{split}
  \FPic{6j-alpha-beta-gamma--adj-adj-adj--loop-clockwise--insert-lambdaj-muk-nul}
  &= \frac{1}{(N^2-1)^3} \scalebox{2.2}{\Bigg[}
  \FPic{6j-alpha-beta-gamma--insert-lambdaj-muk-nul--crossing}
  \\
  & \quad\, 
  - \frac{1}{N} \scalebox{2.2}{\Bigg(}
  \FPic{6j-alpha-beta-gamma--insert-lambdaj-muk-nul--loops-beta-nul}
  + \FPic{6j-alpha-beta-gamma--insert-lambdaj-muk-nul--loops-gamma-lambdaj}
  + \FPic{6j-alpha-beta-gamma--insert-lambdaj-muk-nul--loops-alpha-muk}
  \scalebox{2.2}{\Bigg)}
  \\
  & \quad\,
  + \frac{3}{N^2} \
  \FPic{6j-alpha-beta-gamma--insert-lambdaj-muk-nul--loops-lambdaj-muk-nul}
  - \frac{1}{N^3} \
  \FPic{6j-alpha-beta-gamma--insert-lambdaj-muk-nul--loops-lambdaj-muk-nul--clockwise}
  \scalebox{2.2}{\Bigg]}\;.
\end{split}
\end{equation}

Here the first term needs to be reduced using $6j$ symbols,
\begin{equation}
  \FPic{6j-alpha-beta-gamma--insert-lambdaj-muk-nul--crossing}
  = \FPic{6j-alpha-beta-gamma--insert-lambdaj-muk-nul--uncrossing}
  = \sum_{\sigma} d_\sigma
  \FPic{6j-alpha-beta-gamma--insert-lambdaj-muk-nul--uncrossing--insert-sigma}\nonumber
\end{equation}
\begin{eqnarray}
  \phantom{space bitte...} &=& \sum_{\sigma} d_\sigma \, 
  \FPic{6j-alpha-lambdaj-fund--fundSTAR-beta-sigmaSTAR}
  \FPic{6j-fund-muk-gamma--betaSTAR-fund-sigma}
  \FPic{6j-alpha-sigma-gamma-nul--fund-fund}\\
  &=& \sum_{\sigma} d_\sigma \, 
  \FPic{6j-alpha-lambdaj-fund--fundSTAR-beta-sigmaSTAR}
  \FPic{6j-fund-muk-gamma--betaSTAR-fund-sigma}
  \FPic{6j-fund-nul-alpha--gammaSTAR-fund-sigma}\;,\nonumber
\end{eqnarray}
where a minus sign next to a vertex indicates that the lines are
connected to this vertex in opposite order, see
\Eqref{eq:signed-vertex}.  The expressions calculated here are
assembled in \Eqref{eq:f} and \Eqref{eq:d} for the antisymmetric and
symmetric vertices, respectively.

\section{Vertex normalizations leading to non-trivial $3j$ symbols}
\label{app:restore-3j}

All explicit formulae for Wigner $6j$ symbols in this article, in 
particular the results in \secref{sec:closed-form-expressions}, are valid for
vertices normalized such that all non-vanishing $3j$ symbols are equal
to $1$.
While this normalization is convenient, it differs from normalizations
typically applied in the context of QCD.
We therefore here give a simple rule for how
to transform any of our $6j$ symbols when changing the normalization
of any $3j$ symbol.

Assume we have calculated the $6j$ symbol
\begin{equation}
  \FPic{6j-with-alpha-beta-gamma--Vbullet} \, , 
\end{equation}
whereby we chose the normalization
\begin{equation}
  \FPic{3j-alpha-beta-gamma} = 1 \, .
\end{equation}
If we prefer this $3j$ symbol to be equal to $C\neq 1$ we define a vertex
\begin{equation}
  \FPic{Vertex-betaOut-gammaIn-alphaIn--Vsquare}
  = \sqrt{C} \ \FPic{Vertex-betaOut-gammaIn-alphaIn} \, ,
\end{equation} 
which then satisfies
\begin{equation}
  \FPic{3j-alpha-beta-gamma--Vsquare}
  = C \ \FPic{3j-alpha-beta-gamma} 
  = C \, .
\end{equation}
Consequently, 
\begin{equation}
  \FPic{6j-with-alpha-beta-gamma--Vsquare}
  = \sqrt{C} \ \FPic{6j-with-alpha-beta-gamma--Vbullet} \, .
\end{equation}
\textit{In short:} For each vertex whose $3j$ symbol you normalize to a number
$\neq 1$ multiply our $6j$ symbol by the square root of the value of
your $3j$ symbol in order to obtain the value of the $6j$ symbol with
your normalization convention.

\bibliography{refs}

\end{document}